**Composition-Driven High-Entropy Alloys with Enhanced Magnetocaloric Properties**

*Nishant Tiwari* [a, b]*, Juan Rafael Gomez Quispe* [c]*, Noorbasha Bhavani Sai*[d]*, Saikat Talapatra*[e]*, Pedro Alves Da Silva Autreto**[c]*, Varun Chaudhary**[b] *and Chandra Sekhar Tiwary**[a, d]

Nishant Tiwari, Chandra Sekhar Tiwary

Department of Metallurgical and Materials Engineering, Indian Institute of Technology Kharagpur, West Bengal 721302, India

E-mail: Chandra.tiwary@metal.iitkgp.ac.in

Nishant Tiwari, Varun Chaudhary

Department of Mechanical Engineering, Chalmers University of Technology, Gothenburg SE-41296, Sweden

Email: varunc@chalmers.se

Juan Rafael Gomez Quispe, Pedro Alves Da Silva Autreto

Center for Natural and Human Sciences (CCNH), Federal University of ABC Rua Santa Adélia 166, Santo André 09210-170, Brazil

Email: pedro.autreto@uafbc.edu.br

Noorbasha Bhavani Sai

School of Nano Science and Technology, Indian Institute of Technology, Kharagpur, West Bengal 721302, India

Saikat Talapatra

School of Physics and Applied Physics, Southern Illinois University, Carbondale, IL 62901, USA

**Abstract**

High-entropy alloys (HEAs) are promising magnetocaloric materials with tunable operating temperature conditions using compositional modifications. Here, we combine experiments and first principles-based spin modelling to engineer magnetocaloric response in single-phase cubic HEAs consisting of earth-abundant elements such as Fe, Ni, Co, Cr, and Cu. An equiatomic ($Fe_{20}Ni_{20}Co_{20}Cr_{20}Cu_{20}$-E-HEA) and a Fe/Co-rich non-equiatomic ($Fe_{34}Ni_{17.7}Co_{24.8}Cr_{15.2}Cu_{8.3}$-NE-HEA) show a continuous ferromagnetic-to-paramagnetic transition with Curie temperature ($T_C$) $\approx 110$ K (E-HEA) and $T_C \approx 420$ K (NE-HEA). Under the 1.6 Tesla magnetic field, the investigated alloy shows the entropy change ($|\Delta S_M|_{max}$) $\approx 1.24$ J/kg-K with relative cooling power (RCP) ≈ 92 J/kg due to a broader effective cooling span. Density functional theory simulations reveal that reducing Cu enhances the spin-polarized Fe/Co/Ni-3d weight near $E_F$, consistent with stronger ferromagnetic exchange. Exchange couplings $J_{ij}$ mapped onto a classical Heisenberg model and solved by atomistic Monte Carlo to theoretically predict $T_C$ of both the investigated alloys. A controlled theoretical Cu sweep in equimolar Fe-Ni-Co-Cr further confirms that increasing Cu monotonically dilutes the magnetic sublattice and lowers $T_C$, providing a quantitative design guideline to tune magnetocaloric operating temperatures in transition-metal HEAs.

## 1. Introduction

Magnetic refrigeration is regarded as a sustainable alternative to conventional vapor-compression cooling technologies, and the development of advanced magnetocaloric materials (MCM) is therefore central to the realization of practical refrigeration devices [1]. By utilizing the magnetocaloric effect (MCE), magnetic refrigeration units can achieve a higher efficiency while removing harmful greenhouse gases, which are currently used in conventional vapor compression refrigeration with a solid material known as MCM [2–4]. Traditionally, rare-earth elements such as Gd and La-based alloys have dominated MCE research because of their large magnetic entropy change  and near room transition temperatures [5–8]. Nevertheless, the high cost, limited availability, and susceptibility of rare-earth-based materials to supply-chain disruptions have intensified the search for rare-earth-free alternatives. In this context, Heusler alloys (Ni-Mn-X (X = Ga, Sn, and In)) have emerged as promising rare-earth-free magnetocaloric materials [9–15]; however, their practical applicability is often constrained by poor mechanical properties and large thermal hysteresis losses [16]. Despite these limitations, several studies have demonstrated that magnetocaloric performance and operating temperature ranges can be effectively improved through compositional tuning and targeted alloying approaches [17–23]. In contrast, conventional binary and ternary rare-earth-based systems provide relatively limited compositional flexibility, thereby restricting further optimization and the incorporation of additional alloying elements.

In this context, HEAs have recently attracted considerable interest for magnetocaloric applications. The compositional flexibility and entropy-stabilized phases of HEAs offer a promising pathway to tailor magnetic transitions and enhance MCE-related properties[24]. Unlike conventional alloys, HEAs offer flexibility in adjusting transition temperatures and entropy changes by varying constituent elements, along with improved mechanical properties, which can possibly increase the service life [25]. More recently, the conventional concept of HEAs has expanded towards multiphase systems as well, enthalpy-stabilized materials, and alloys that do not necessarily crystallize as single-phase solid solutions[26]. As a result, these materials are now more broadly categorized as compositionally complex alloys or materials [27,28]. Earlier HEAs developed for magnetocaloric applications were primarily based on rare-earth elements or Ge-containing transition-metal-based systems, which often suffer from high cost and limited availability of resources [29]. Despite the promising potential of these materials, establishing reliable composition-property relationships remains essential for the rational design of rare-earth-free magnetocaloric HEAs, particularly in transition-metal-based systems where individual alloying elements can strongly influence the magnetic behavior.

In the present work, two Fe-Ni-Co-Cr-Cu-based HEAs (equiatomic and non-equiatomic) were synthesized and investigated as rare-earth-free MCM. The optimized composition of the Fe-Ni-Co-Cr-Cu-based HEAs system was determined using FactSage thermochemical software. The optimization process integrates Gibbs free energy minimization with a Pareto-based methodology, enabling the simultaneous consideration of multiple constraints. Thermodynamic properties of the system were evaluated using the integrated databases available within FactSage[30]. A series of constraints was imposed during the optimization process, including liquidus temperature, homogenization temperature, phase constitution, pressure, entropy, and enthalpy. The constraints imposed include a single-phase FCC structure, a homogenization temperature of 800 °C, and a configurational entropy greater than 1.5R (R being the universal gas constant). Both alloys crystallize in an FCC structure and exhibit a ferromagnetic-to-paramagnetic transition. As depicted in **Figure 1(a),** the Fe- and Co-rich non-equiatomic composition (NE-HEA) shows a higher Curie temperature (~420 K, *yellow*), a higher magnetization at low temperatures, and a substantial magnetocaloric properties near room temperature, whereas the equiatomic (E-HEA) alloy presents a stronger magnetocaloric response at lower temperatures (~110 K, *pink*). These experimental results indicate that controlled variations in composition provide a practical route to engineer magnetocaloric performance in Fe-Ni-Co-Cr-Cu HEAs. In parallel, we develop a microscopic understanding of how Cu content and chemical disorder modify the electronic structure, exchange interactions, and Curie temperature in this alloy family.

To this end, we combine density functional theory (DFT) with atomistic Monte Carlo (MC) simulations. We use DFT to compute element-resolved projected densities of states and to extract Heisenberg exchange parameters for equiatomic and non-equiatomic Fe-Ni-Co-Cr-Cu alloys with varying Cu content, and then employ large-scale MC simulations to obtain finite-temperature magnetization curves and Curie temperatures. The multiscale approach provides a theoretical framework to rationalize experimental observations and to guide the design of optimized Fe-Ni-Co-Cr-Cu HEAs for magnetic and magnetocaloric applications. In addition to its relevance for magnetic refrigeration technologies, this approach also offers a versatile strategy for the design of advanced MCM that shows higher efficiency and sustainability.

## 2. Results and Discussion

### 2.1 Structural characterization

The structural characteristics of the investigated alloys were examined using various techniques. **Figures 1(b-h)** present the XRD (X-Ray diffraction) patterns, SEM (Scanning Electron Microscopy) images obtained in both SE and BSE (secondary and backscattered electron) modes, along with the HR-TEM (High Resolution-Transmission Electron Microscopy) images of the studied alloys. Prior to characterization, the samples were mechanically polished to remove any oxide layer formed because of the high Fe content of the alloys. The XRD results, shown in Figures 1(b) and (c), reveal that both investigated alloys exhibit a cubic crystal structure. Furthermore, the diffraction peaks of the E-HEA and NE-HEA samples were successfully indexed to a face-centered cubic phase. The value of the lattice parameter determined from the XRD analysis was found to be 3.72 Å for E-HEA and 3.65 Å for NE-HEA. Detailed microstructural characterization of both alloys was performed using SEM. The equiatomic HEA (E-HEA) reveals a predominantly single-phase matrix along with a distinct Cu-rich secondary phase, as evidenced by the SE and BSE images mentioned in Figure 1(d) and **Figure S1**, respectively. Phase contrast in the BSE images clearly differentiates the Cu-enriched regions from the matrix due to compositional variations, as shown in **Figure S1(a)**. X-ray diffraction analysis indicates that both the matrix and the Cu-rich phase crystallize in an FCC structure, consistent with the SEM observations. Elemental mapping and energy-dispersive spectroscopy (EDS) analysis (Figure S1(b-f)) further confirm the homogeneous distribution of Fe, Ni, Co, Cr, and Cu within the matrix, along with localized Cu segregation. Additionally, point EDS measurements (Figure S1(g) and (h)) were conducted to quantitatively verify the compositional differences between the matrix and the Cu-rich phase, corroborating the phase segregation behavior. Similarly, SEM analysis of the NE-HEA in both SE and BSE modes, as shown in Figure 1(e) and **Figure S2**, reveals a similar microstructural morphology. However, in contrast to the equiatomic alloy, the NE-HEA exhibits a noticeably lower fraction of Cu-rich precipitates, which can be attributed to its reduced Cu content, as shown in **Figure S2(a)**. The diminished compositional contrast in the BSE images further supports the decreased extent of Cu segregation. Elemental mapping and EDS analysis together (Figure S2(b-f)) confirm the homogeneous distribution of Fe, Ni, Co, Cr, and Cu within the matrix, along with limited localized Cu enrichment. Additionally, point EDS measurements were carried out to quantitatively determine the compositions of the matrix and Cu-rich regions, substantiating the reduced secondary phase formation in the NE-HEA. The formation of a Cu-rich phase in Fe-Co-Ni-Cr-Cu HEAs has also been reported in earlier studies and is supported by thermodynamic predictions [31]. Further HR-TEM analysis was carried out for the NE-HEA alloy, as shown by the STEM image in Figure 1(f). To obtain detailed structural information, FFT (Fast Fourier transformation) and IFFT (Inverse Fast Fourier transformation) analyses were performed on the STEM image, confirming the FCC crystal structure, as presented in Figure 1(g). Additionally, line profile analysis derived from the IFFT image reveals the presence of dislocations, highlighted by red markers in Figure 1(h), indicating the existence of lattice imperfections in the NE-HEA alloy.

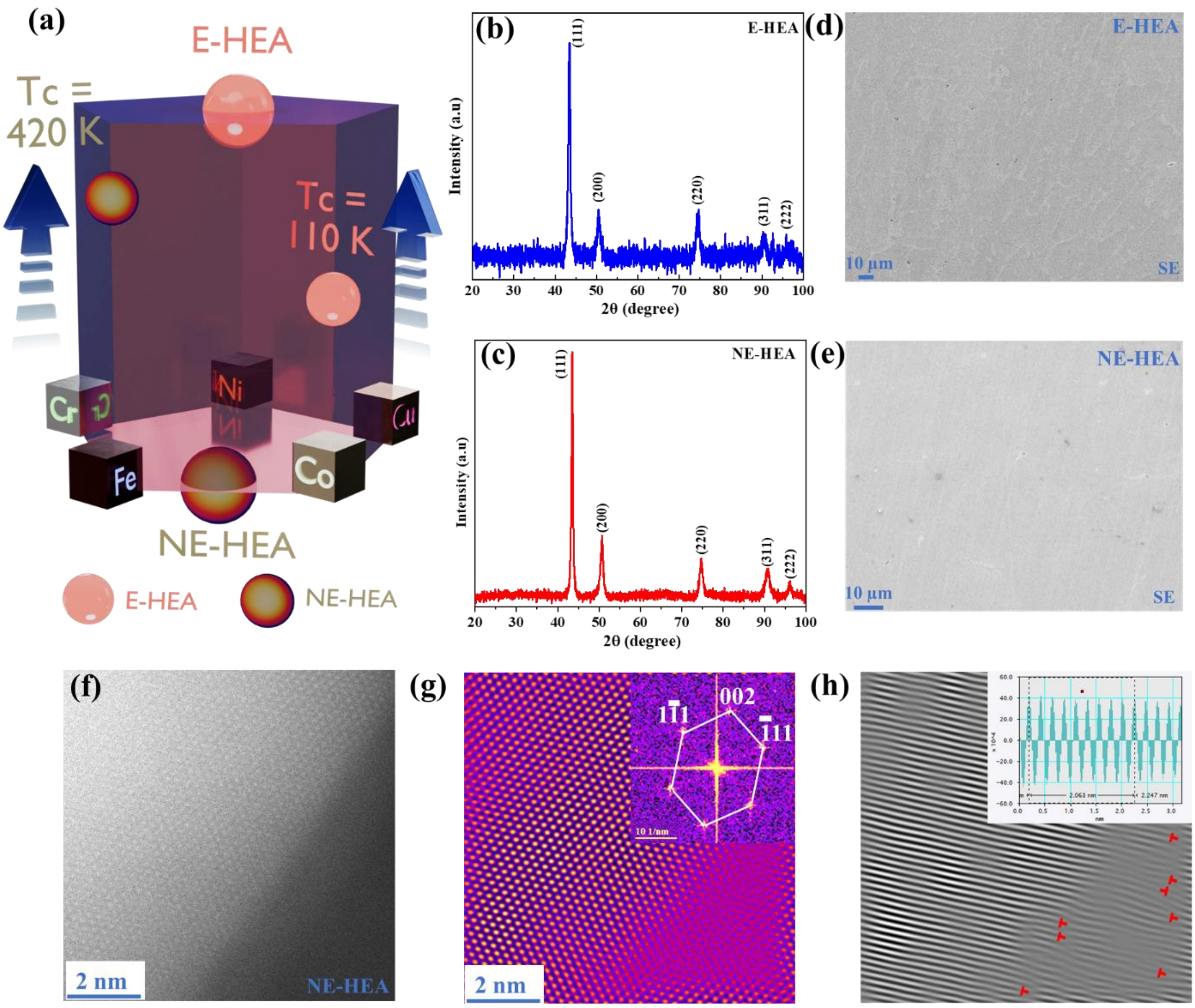

**Figure 1. (a)** Schematic showing both the E-HEA *(pink and top)* and NE-HEA *(yellow and bottom)* constituent elements and how their curie temperature (*Right for E-HEA, Left for NE-HEA)* can vary while moving from E-HEA *(110 K)* to NE-HEA *(420 K)*. **(b)** E-HEA XRD behavior. **(c)** NE-HEA XRD behavior **(d)** SEM Micrograph of E-HEA in SE mode **(e)** SEM Micrograph of NE-HEA in SE mode **(f)** STEM image of NE-HEA **(g)** Inverse FFT (Fast Fourier Transformation), and FFT of STEM micrograph of NE-HEA (inset) **(h)** Line profile of STEM image along (111) plane along with the marked dislocation presence (red).

### 2.2 Magnetocaloric Behaviour

The temperature-dependent magnetization (M-T) for the equiatomic (E-HEA) $Fe_{20}Ni_{20}Co_{20}Cr_{20}Cu_{20}$ and the non-equiatomic (NE-HEA) $Fe_{34}Ni_{17.7}Co_{24.8}Cr_{15.2}Cu_{8.3}$ alloys is shown in **Figure 2(a)** and **(d)**, respectively. Both alloys show a ferromagnetic-paramagnetic (FM-PM) transition at an applied magnetic field of 0.1 T (Tesla), although the magnetic response is strongly influenced by compositional variation. For the E-HEA alloy, the magnetization is relatively high at low temperatures due to cooperative ferromagnetic interactions among Fe, Ni, and Co atoms, as shown in Figure 2(a). However, the presence of non-magnetic Cu and antiferromagnetically coupled Cr reduces the overall magnetization and suppresses the Curie temperature ($T_C$). The magnetization decreases smoothly with increasing temperature, exhibiting a typical sigmoidal drop characteristic of a second-order FM-PM transition. Above $T_C$, the alloy displays paramagnetic behaviour with only weak field-induced magnetization. The broadness of the transition can be attributed to local chemical disorder, which is intrinsic to high-entropy alloys. To clearly determine the $T_C$, the first derivation of the M-T curve as a function of temperature (dM/dT vs T (K)) was also plotted. The resulting derivative curve exhibits a distinct extremum corresponding to a $T_C$ value around 110 K, as shown in Figure 2(b).

In contrast, the NE-HEA alloy displays a distinctly different M-T response, as shown in Figure 2(d). The higher Fe and Co concentrations enhance the saturation magnetization at low temperatures, reflecting stronger ferromagnetic exchange interactions compared to the equiatomic case. Additionally, the reduced Cu content decreases the extent of magnetic dilution, while a lower Cr concentration diminishes antiferromagnetic frustration. Together, these effects shift the Curie temperature upward and increase the overall magnetic moment. Interestingly, the M-T curves obtained during heating and cooling (red and blue) show a small degree of thermal hysteresis near $T_C$, suggesting the presence of magneto-structural coupling effects that slightly alter the transition pathway. As is evident from the magnetization data, the magnetic transition exhibits characteristics of a second-order phase transition, for which the determination of the $T_C$ using a simple derivative is not sufficient. In second-order transitions, the magnetization changes gradually over a broad temperature range rather than showing a sharp discontinuity, leading to uncertainty in locating $T_C$ solely from dM/dT analysis. To obtain a more reliable estimation of $T_C$, Arrott plots ($M^2$ vs H/M) were therefore constructed. According to the Banerjee criterion and mean-field theory, the $T_C$ corresponds to the isotherm that passes closest to the origin in the Arrott plot [32–34]. Using this approach, the $T_C$ was determined to be approximately 420 K, as shown in Figure 2(e), providing a more robust and physically meaningful evaluation of the magnetic transition temperature.

Comparing the two alloys, it is evident that compositional tuning strongly governs the balance between ferromagnetic and competing non-magnetic/antiferromagnetic interactions. The E-HEA alloy, with equal fractions of constituent elements, shows suppressed magnetization and a broadened second-order transition. In contrast, the Fe- and Co-rich NE-HEA alloy exhibits enhanced ferromagnetism, a higher $T_C$, and stronger magnetic stability against thermal fluctuations. These observations highlight the importance of strategic compositional design in high-entropy alloys based on transition elements, where controlling the balance between magnetic and non-magnetic elements directly influences their suitability for magnetocaloric applications.

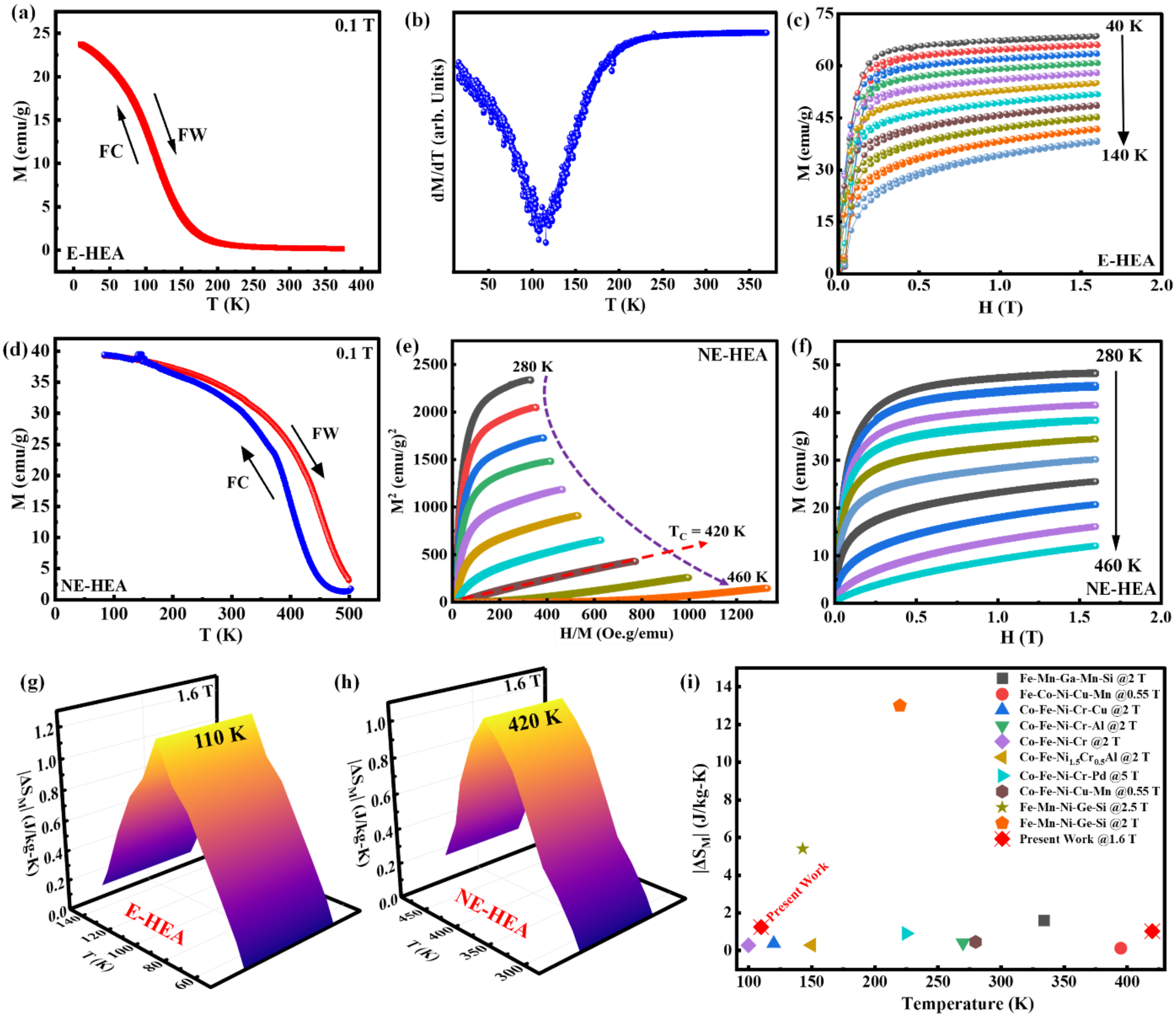

**Figure 2. (a)** Magnetization curve with respect to temperature (M-T) behavior in Field warmed (FW) and Field Cooling (FC) mode of E-HEA at 0.1 T. **(b)** dM/dT curve of M-T behavior of E-HEA **(c)** M-H curves of E-HEA in the range of 40 K-140 K at the interval of 10 K **(d)** M-T

behavior in Field warmed (FW) and Field Cooling (FC) mode of NE-HEA at 0.1 T. **(e)** Arrott plots showing $T_C$ of NE-HEA around 420 K **(f)** M-H curves of NE-HEA in the range of 280 K-460 K at the interval of 20 K. **(g)** Value of $|\Delta S_M|$ across M-(H) curves of the E-HEA at 1.6 T. **(h)**Value of $|\Delta S_M|$ across M-(H) curves of the NE-HEA at 1.6 T. **(i)** Comparison of the investigated alloy along with other standard magnetocaloric materials [35–44].

To evaluate the magnetocaloric performance of the investigated alloys, M-H measurements were carried out across the respective transition temperature regions, as illustrated in Figures 2(c) and 2(f). The isothermal M-H curves were recorded at temperature intervals of 10–20 K within the ranges of 40-140 K for E-HEA and 280-460 K for NE-HEA. Furthermore, the magnetic entropy change ($|\Delta S_M|$) and relative cooling power (RCP) associated with these magnetic transitions were determined through numerical integration of Maxwell's relations, as expressed in **Equations (1)** and **(2)**, respectively.

$$\Delta S_M\left(\mathrm{T}, \mathrm{H_{applied}}\right) = \mu_0 \int_0^{H_{applied}} \left(\frac{\partial \mathrm{M}}{\partial \mathrm{T}}\right)_{p,H} dH \quad \textbf{(1)}$$

$$\mathrm{RCP} = \Delta \mathrm{S_M^{Max}} \times (\Delta \mathrm{T_{FWHM}}) \quad \textbf{(2)}$$

As illustrated in Figure 2(g), the E-HEA alloy exhibits a maximum magnetic entropy change ($|\Delta \mathrm{S_M^{Max}}|$) for E-HEA is 1.24 J $kg^{-1}$ $K^{-1}$at approximately 110 K under an applied magnetic field of 1.6 T. In addition, the corresponding relative cooling power (RCP) reaches 75.2 J $kg^{-1}$, indicating effective heat transfer capability between the cold and hot reservoirs in magnetic refrigeration applications. In comparison, the NE-HEA alloy demonstrates a slightly reduced $|\Delta \mathrm{S_M^{Max}}|$ of 1.02 J $kg^{-1}$ $K^{-1}$, while exhibiting a higher RCP value of 91.8 J $kg^{-1}$ at around 420 K, as shown in Figure 2(h) (*red curve*). Furthermore, the magnetocaloric performance of the present alloys was systematically compared with previously reported transition-metal-based HEAs, as shown in Figure 2(i). This comparison reveals that the magnetic entropy change and effective operating temperature range achieved in the current study are comparable to those of established transition-metal-based magnetocaloric systems, thereby underscoring the competitiveness and practical relevance of the investigated HEAs as alternative magnetocaloric materials.

## 2.3 Theoretical results

**Figure 3** compares the atomic structures and element-resolved spin-polarized projected (into 3d orbitals) density of states (PDOS) of the equiatomic (E-HEA, $Fe_{0.20}Ni_{0.20}Co_{0.20}Cr_{0.20}Cu_{0.20}$) and non-equiatomic (NE-HEA, $Fe_{0.34}Ni_{0.177}Co_{0.248}Cr_{0.152}Cu_{0.083}$) alloys. Figure 3(a) and (b) show representative chemically disordered FCC supercells used in the DFT calculations, where all five elements randomly occupy the lattice sites according to the target compositions. This modelling strategy is consistent with previous theoretical studies of transition-metal HEAs, which have demonstrated that random or special quasi-random structures capture well the local chemical environments and reproduce experimental phase stability and magnetic trends in multi-principal alloys such as Co-Cr-Cu-Fe-Mn-Ni and related systems [45–47]. In particular, the predominance of the FCC structure in both E-HEA and NE-HEA is consistent with XRD indexing in Fe-Co-Ni-Cr-Cu alloys. However, bulk alloys in this system frequently exhibit Cu segregation, leading to a Cu-lean FCC matrix and a Cu-rich FCC phase/clusters; because both constituents share the FCC structure, their diffraction peaks can overlap, so XRD may appear as a single FCC phase unless complemented by microstructural/chemical analyses [48].

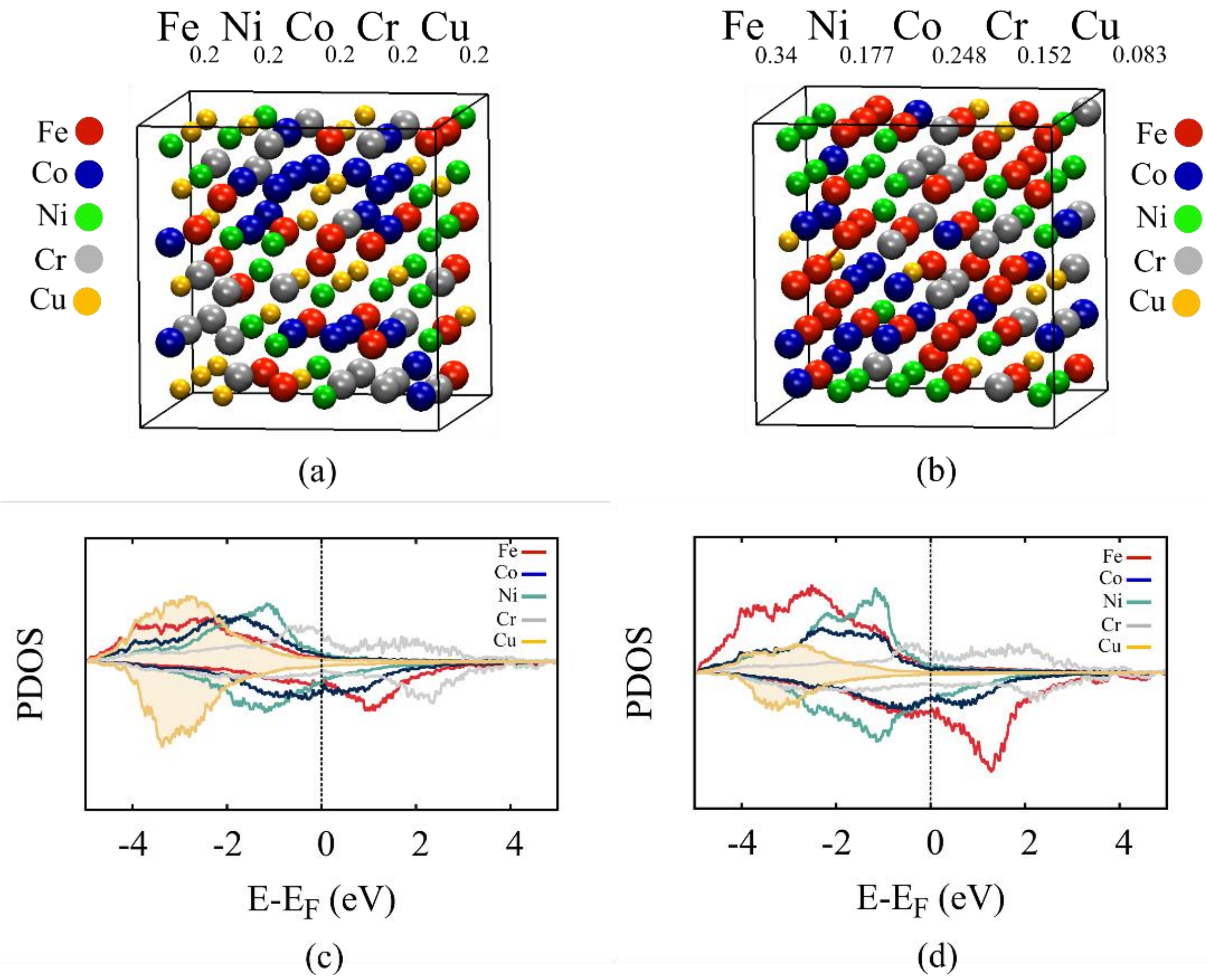

**Figure 3.** Atomic structures and electronic properties of the E-HEA and NE-HEA Fe-Ni-Co-Cr-Cu high-entropy alloys. **(a, b)** Supercell models used in the DFT calculations for the E-HEA ($Fe_{0.20}Ni_{0.20}Co_{0.20}Cr_{0.20}Cu_{0.20}$) and NE-HEA ($Fe_{0.34}Ni_{0.177}Co_{0.24}Cr_{0.152}Cu_{0.083}$), respectively; atoms are colored according to the legend. **(c, d)** Element-resolved spin-polarized projected (3d) density of states (PDOS) for E-HEA and NE-HEA, respectively, as a function of energy relative to the Fermi level $E_F$; positive (negative) values denote majority (minority) spin states.

The PDOS in Figure 3(c) and (d) correspond to the E-HEA and NE-HEA, respectively. In both cases the finite DOS at the Fermi level confirms the metallic character of these alloys, with no gap opening at the Fermi energy. For the E-HEA alloy Figure 3(c), the 3d states of Fe, Co and Ni dominate in the vicinity of the $E_F$, while Cu-3d states lie mostly at lower energies but still retain a small, slightly spin-asymmetric contribution, consistent with a very weak local moment on Cu [49]. In contrast, the NE-HEA Figure 3(d) exhibits a strongly reduced Cu-3d weight and a marked increase of the Fe-, Co- and Ni-3d contributions, together with a clearer spin imbalance in these channels. This comparison shows that elemental redistribution from E-HEA to NE-HEA reshapes the 3d manifold near the Fermi level: lowering the Cu content enhances the Fe/Co/Ni-derived states that carry the magnetism, which is expected to strengthen the ferromagnetic exchange and, consequently, to increase the Curie temperature of the NE-HEA.

**Figure 4** presents the atomistic simulation results for the M-T curves in both the E-HEA and NE-HEA alloy. All calculations were performed in a cubic simulation cell with a side length of 15 nm to minimize finite-size effects. As shown in **Figure 4(a),** the elemental redistribution from an equimolar to a non-equimolar composition enriched in Fe, Co, and Ni due to that a pronounced increase in the value of Tc, from $393.3 \pm 9.6$K for E-HEA to $648.3 \pm 9.5$K for NE-HEA. **Figures 4(b-d)** and **4(e-g)** display representative snapshots of the atomic magnetic moments projected onto the z axis for E-HEA and NE-HEA, respectively, at $T = 100$, 300, and 500 K. At 100 K, NE-HEA exhibits an almost fully saturated configuration with moments strongly aligned along z, evidencing a robust ferromagnetic coupling. In contrast, E-HEA already shows regions where the local moments deviate from the z direction, indicating a weaker effective ferromagnetic interaction. At 300 K, both alloys display a reduced magnetization due to enhanced thermal fluctuations competing with the exchange interaction, although ferromagnetically aligned domains remain dominant in both cases. At 500 K, E-HEA has clearly transformed into a paramagnetic state: the local moments are essentially random, and the macroscopic magnetization vanishes. NE-HEA, however, still retains sizeable regions with parallel alignment at 500 K, consistent with its substantially higher $T_C$ and stronger ferromagnetic stability compared to the E-HEA alloy.

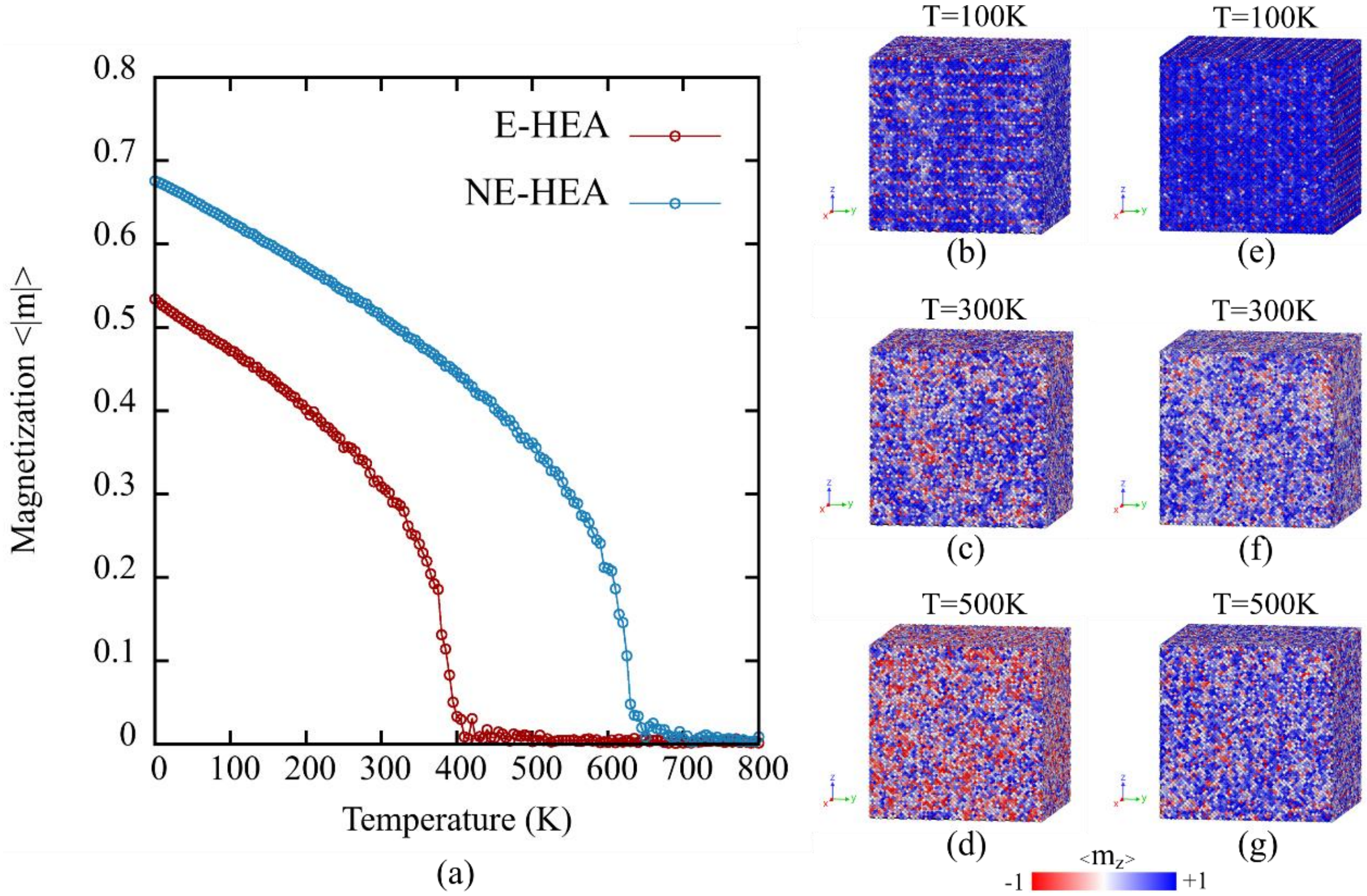

**Figure 4.** Atomistic Monte Carlo Simulations results for the equiatomic (E-HEA) and Non-equiatomic (NE-HEA) Fe-Ni-Co-Cr-Cu high-entropy alloys. **(a)** Temperature dependence of the normalized magnetization ⟨|m|⟩ obtained from VAMPIRE simulations, showing a substantially higher Curie temperature for NE-HEA (blue) compared to E-HEA (red). **(b-d)** Representative spin configurations of E-HEA at T = 100 K, 300 K, and 500 K, respectively. **(e-g)** Corresponding spin configurations of NE-HEA at the same temperatures. The color scale encodes the local z-component of the normalized magnetic moment $\langle m_z \rangle$, from −1 (red) to +1 (blue).

A comparison between the calculated and experimentally measured Curie temperatures reveals a discrepancy of approximately ~230-280 K, as shown in **Figure 4(a)** and **Figure 2(a, d),** respectively. Nevertheless, the relative difference in $T_C$ between the two investigated alloys remains consistent in both the theoretical predictions and experimental measurements. The discrepancy mainly reflects the mapping and associated approximations: the exchange parameters $J_{ij}$ are extracted from first-principles calculations at 0 K and treated as temperature-independent; the classical Heisenberg model assumes rigid magnetic moments; and the finite real-space cutoff in $J_{ij}$ may neglect longer-range exchange contributions [50,51]. Furthermore, microstructural defects such as dislocations, lattice strain, and chemical disorder introduce local variations in atomic environments and exchange interactions that are not captured in idealized theoretical models. In our structural characterization, the presence Cu rich precipitates and dislocations in both the E-HEA and NE-HEA (highlighted in **Figure 1(c), (f), and (j)**) is a significant microstructural disorder. Such real-world imperfections are expected to disrupt long-range magnetic order and thus lower the transition temperature relative to computations performed under idealized assumptions, contributing to the observed experimental - theoretical $T_C$ discrepancy [52,53].

Additionally, Huang et al. [54] reported Tc = 251 K for the equiatomic FCC CrFeCoNiCu alloy, which is lower than our E-HEA value (Tc = 393.3 ± 9.6 K). Such differences are expected due to distinct methodological assumptions: CPA/DLM combined with a mean-field Tc mapping [54], versus our supercell DFT-derived $J_{ij}$ combined with explicit Monte Carlo sampling. Differences in disorder averaging, Tc estimator (mean-field versus Monte Carlo), and exchange parametrization/range (including the real-space truncation of $J_{ij}$) can shift absolute Tc values even when composition-dependent trends remain consistent.

## 3. Discussion

To highlight the intrinsic effect of Cu on the ferromagnetic coupling in the homogeneous solid-solution limit, four disordered FCC HEA models with $Cu_x$ = 7.4, 14.8, 29.6, and 37.0 at. % were constructed while keeping Fe, Ni, Co, and Cr equimolar. The temperature-dependent magnetization (M-T) in Figure 5(a) shows a monotonic trend: The Tc decreases systematically with the Cu content increases; accordingly, the highest Tc values within the investigated set correspond to the lowest Cu concentrations (7.4% and 14.8%). This behaviour is consistent with magnetic dilution, since Cu is (nearly) non-magnetic and its substitution reduces the number of Fe/Co/Ni magnetic neighbours and weakens the effective ferromagnetic exchange network. Consistently, the PDOS in Figure 5(f-i) exhibits a progressive depletion of Fe-, Co-, and Ni-3d states near EF as Cu increases, indicating a reduced spin-polarized 3d manifold that correlates with weaker ferromagnetic stability and lower Tc.

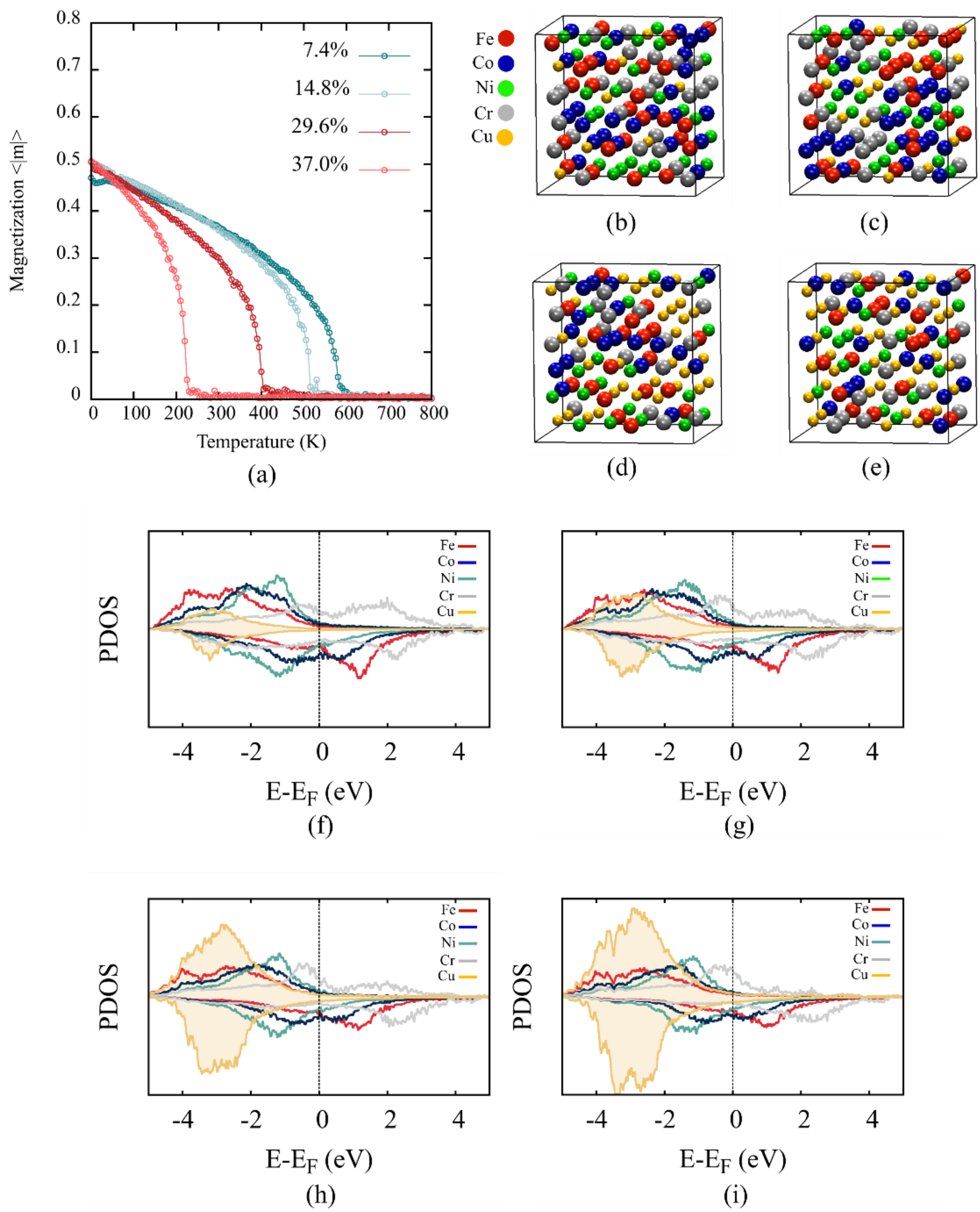

**Figure 5.** Effect of Cu content on the electronic structure and magnetization of the Fe-Ni-Co-Cr-Cu high-entropy alloy. **(a)** Temperature dependence of the normalized magnetization ⟨|m|⟩ for four Cu concentrations (7.4, 14.8, 29.6, and 37.0 at. % Cu) obtained from atomistic Monte Carlo simulations, showing a systematic reduction of the Curie temperature with increasing Cu content. **(b-e)** Representative supercell configurations used in the DFT calculations for the corresponding Cu concentrations; atoms are color-coded by chemical species. **(f-i)** Element-resolved spin-polarized projected density of states (PDOS) for each Cu concentration as a function of energy relative to the Fermi level $E_F$; positive (negative) values correspond to majority (minority) spin states, illustrating the progressive weakening of the 3d magnetic states with increasing Cu substitution.

This overall picture complements and also consistent with the experimental findings of Chaudhary et al. on Fe-Ni-Co-Cr-$Cu_x$ high-entropy alloys[48]. In their study, the introduction of non-magnetic Cu into Fe-Ni-Co-Cr leads to the formation of Cu-rich regions and a Cu-lean matrix enriched in Fe and Co, which enhances the effective exchange interactions and increases the Curie temperature. In our case, we achieve an analogous effect by explicitly tuning the bulk composition from E-HEA to NE-HEA and by systematically varying the Cu content at fixed Fe-Ni-Co-Cr ratios: the PDOS analysis shows that reducing Cu and enriching Fe/Co/Ni increases the 3d spin polarization at the Fermi level, while Monte Carlo simulations based on DFT-derived exchange couplings yield a substantial enhancement of $T_C$. Taken together, these results provide a quantitative electronic-structure and spin-model framework that supports the mechanism inferred by Chaudhary et al.[48], namely that elemental redistribution and Cu partitioning concentrate Fe/Co/Ni in the magnetically active states and thereby strengthen ferromagnetic exchange in Fe-Ni-Co-Cr-Cu high-entropy alloys. The theoretical predictions were further validated by comparison with the present experimental results, as shown in **Figure 6**. A consistent trend is observed in both approaches, wherein the relevant magnetic properties increase with decreasing Cu content from the NE-HEA (8.3 at. %) to the E-HEA (20 at. %). However, the NE-HEA composition also involves variations in other alloying elements, particularly Fe, which can significantly influence $T_C$. The addition of Fe is well known to enhance ferromagnetic exchange interactions due to its large magnetic moment and strong 3d-3d coupling, thereby increasing $T_C$. Consequently, the observed increase in $T_C$ for the NE-HEA cannot be attributed solely to Cu content, but rather arises from the combined effects of compositional complexity and Fe-induced enhancement of magnetic exchange interactions[55,56].

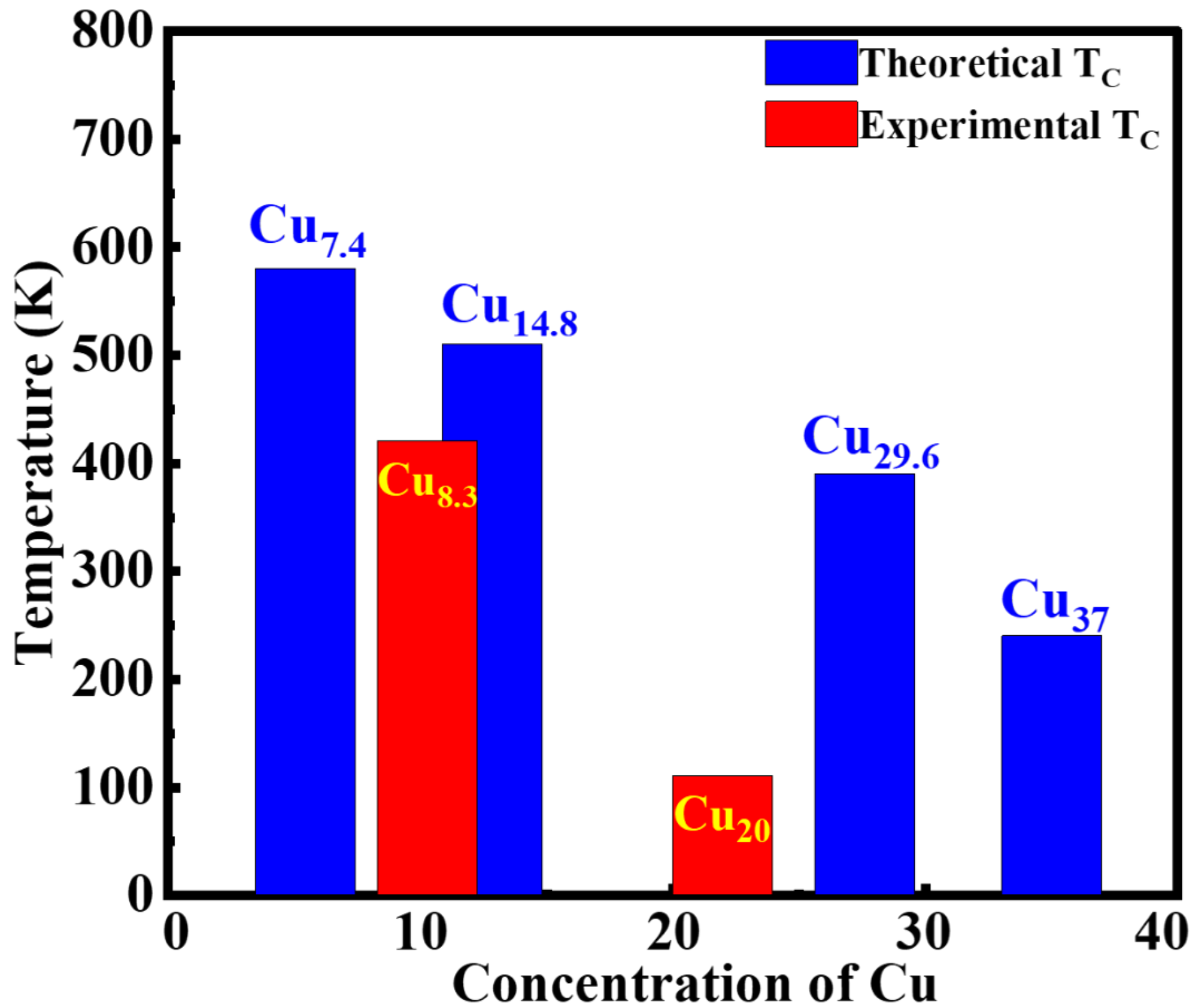

**Figure 6.** Effect of Cu content in both theoretical and experimental compositions on the Curie Temperature ($T_C$) of the Fe-Ni-Co-based HEAs, which are studied in the present work.

## 4. Conclusions

In this work, we used an integrated approach of DFT and experimental to evaluate the magnetocaloric properties of equiatomic (E-HEA) and non-equiatomic (NE-HEA) Fe-Ni-Co-based high-entropy alloys. Structural characterization confirmed that both alloys crystallize in the FCC structure, and magnetic measurements were used to quantify their magnetocaloric response. The study highlights the role of Cu, a diamagnetic element, as an effective compositional tuning parameter for functional properties, enabling Curie temperature shift and the associated operating temperature window. The E-HEA exhibits a higher peak magnetic entropy change ($|\Delta S_M| \approx 1.24$ J/kg-K) with a RCP of $\sim 75.2$ J/kg, whereas the NE-HEA shows a slightly lower $|\Delta S_M|$ peak ($\approx 1.02$ J/kg-K) but an enhanced relative cooling power of $\sim 91.8$ J/ kg due to its wider effective cooling span.

To rationalize these trends, we combined the experiments with first-principles-based modelling to elucidate how composition-driven changes influence the magnetic behaviour of Fe-Ni-Co-Cr-Cu high-entropy alloys. For the E-HEA and NE-HEA compositions, we find that both systems remain metallic; however, the Fe- and Co-rich NE-HEA displays enhanced spin polarization of Fe-, Co-, and Ni-3d states and a strongly reduced Cu-3d contribution at the Fermi level, consistent with a more robust ferromagnetic ground state. Exchange couplings extracted from DFT yield, at finite temperature, a pronounced increase of the theoretical Curie temperature from $393.3 \pm 9.6$ K (E-HEA) to $648.3 \pm 9.5$ K (NE-HEA). Furthermore, by systematically varying the Cu content in an equimolar Fe-Ni-Co-Cr matrix, we show that higher Cu concentrations dilute the magnetic sublattice, suppress the Fe-, Co-, and Ni-3d weight at the Fermi level, and monotonically reduce $T_C$. Overall, these results demonstrate a clear trade-off between peak entropy change and effective cooling span governed by compositional tuning via Cu incorporation, and provide practical guidance for optimizing only transition metal Fe-Ni-Co-Cr-Cu-based high-entropy alloys for magnetocaloric applications.

## 5. Materials and Methods

### 5.1 Experimental Details

Two alloys, each weighing approximately 20 g and having the compositions listed in **Table 1**, were synthesized by arc melting high-purity (99.9%) Fe, Ni, Co, Cr, and Cu elements under an argon atmosphere. To minimize oxidation, particularly due to the elevated Fe concentration, the as-cast samples were encapsulated in evacuated quartz tubes backfilled with argon gas. The prepared alloys were then homogenized through annealing at 1073 K for 7 days, then quenched in ice water.

**Table 1:** Composition (~atomic %) of the high entropy alloys (HEAs) in the present work.

| Samples | Fe | Ni | Co | Cr | Cu |
|---|---|---|---|---|---|
| **(E-HEA)**<br>**Equiatomic- HEA** | 20 | 20 | 20 | 20 | 20 |
| **(NE-HEA)**<br>**Non-Equiatomic- HEA** | 34 | 17.7 | 24.8 | 15.2 | 8.3 |

The crystal structures of the quenched alloys were analyzed using a Bruker D8 diffractometer, Cu-Kα radiation source over a 2θ range of 20°-90°. The microstructural morphology and elemental composition of the samples were investigated using a ZEISS GEMINI 600 field-emission scanning electron microscope (FE-SEM) and JEOL JEM 2100F 300 KV HR-TEM, enabling detailed qualitative as well as quantitative evaluation of the alloys' microstructural characteristics. The magnetic characterizations were assessed using the SQUID (Superconducting Quantum Interference Device) module of an MPMS (Quantum Design, USA) and the VSM (Vibrating Sample Magnetometer) of Lakeshore 8600. The temperature-dependent magnetization (M-T) measurements were carried out under field-cooled cooling (FCC) and field-cooled warming (FCW) conditions using an applied magnetic field of 0.1 Tesla (T). In addition, magnetization as a function of magnetic field (M-H) was measured over a range of 0-1.6 Tesla around phase transition temperatures such as $T_C$ in the present work.

### 5.2 Computational Details

The electronic structure of both the E-HEA and NE-HEA were investigated within spin-polarized density-functional theory (DFT) as implemented in the SIESTA v5.4 software [57,58]. Chemically disordered equiatomic (E-HEA) and non-equiatomic (NE-HEA) compositions were modelled by supercells containing 108 atoms (five chemical species) with cubic lattice vectors $a = b = c = 11.16$ Å, in which Fe, Ni, Co, Cr, and Cu atoms occupy FCC lattice sites at random according to the target concentrations. For each composition, a representative disordered configuration was generated and structurally relaxed. Core electrons were described using norm-conserving pseudopotentials from the PseudoDojo library. Valence electrons were expanded in a numerical atomic-orbital basis set of double-ζ with polarization functions (DZP) [58,59]. A real-space grid cutoff of 300 Ry was employed. Self-consistency was achieved using a Pulay mixing scheme with a density-matrix convergence tolerance of $1 \times 10^{-4}$and up to 5000 self-consistent iterations. Exchange-correlation effects were described within a generalized-gradient approximation (GGA) using the Perdew-Burke-Ernzerhof (PBE) parametrization [60]. Brillouin-zone integrations were performed using Monkhorst-Pack k-point [61] meshes (2x2x2) to ensure convergence of the total energy and magnetic moments. All atomic positions were relaxed until the residual forces were below a small threshold (0.025 eV/Å), and spin-polarized projected densities of states (PDOS) were obtained by projecting the Kohn-Sham states onto atomic orbitals for each chemical species.

Based on the converged DFT solutions, the magnetic exchange interactions were mapped onto a classical Heisenberg model using the TB2J package. TB2J evaluates pairwise exchange coupling parameters $J_{ij}$ from the DFT electronic structure and constructs the effective Hamiltonian[62,63]:

$$H = -\sum_{i \neq j} J_{ij}\, S_i . S_j\ ,$$

where $\mathrm{S}_i$ are classical unit vectors representing the local magnetic moments. A real-space cutoff of 4.0 Å was used, which includes exchange interactions up to second-nearest neighbours in the FCC lattice.

Finite-temperature magnetic properties were then computed using the VAMPIRE atomistic spin-dynamics code [63,64]. For each composition, the Heisenberg Hamiltonian is defined by the DFT-derived $J_{ij}$ was simulated on three-dimensional supercells with periodic boundary conditions. The temperature-dependent magnetization (M-T) curve was obtained by Metropolis Monte Carlo simulation, in which trial spin updates are accepted or rejected according to the Metropolis criteria [63–65]. At each temperature, the system was first equilibrated over enough Monte Carlo steps per spin (50,000 steps for equilibrating and averaging), after which thermal averages of the magnetization were accumulated. The $T_{\mathrm{C}}$ value was estimated from the M-T curve (inflection point) and cross-checked by the vanishing of the spontaneous magnetization within numerical accuracy.

**Data Availability Statement**

The data for the study is made available in the supporting information.

## References

[1] V. Franco, J. S. Blázquez, J. J. Ipus, J. Y. Law, L. M. Moreno-Ramírez, A. Conde, *Prog. Mater. Sci.* **2018**, *93*, 112.

[2] F. Zhang, X. Miao, N. van Dijk, E. Brück, Y. Ren, *Adv. Energy Mater.* **2024**, *14*, DOI 10.1002/aenm.202400369.

[3] B. G. Shen, J. R. Sun, F. X. Hu, H. W. Zhang, Z. H. Cheng, *Advanced Materials* **2009**, *21*, 4545.

[4] T. Gottschall, K. P. Skokov, M. Fries, A. Taubel, I. Radulov, F. Scheibel, D. Benke, S. Riegg, O. Gutfleisch, *Adv. Energy Mater.* **2019**, *9*, DOI 10.1002/aenm.201901322.

[5] V. K. Pecharsky, K. A. Gschneidner, *Giant Magnetocaloric Effect in Gd 5 Si 2 Ge 2*, **1997**.

[6] S. Y. Dan'kov, A. M. Tishin, V. K. Pecharsky, K. A. Gschneidner, *Magnetic Phase Transitions and the Magnetothermal Properties of Gadolinium*, **1998**.

[7] J. Liu, C. He, M. X. Zhang, A. R. Yan, *Acta Mater.* **2016**, *118*, 44.

[8] Y. Zhang, Y. Na, W. Hao, T. Gottschall, L. Li, *Adv. Funct. Mater.* **2024**, *34*, DOI 10.1002/adfm.202409061.

[9] K. Mandal, D. Pal, N. Scheerbaum, J. Lyubina, O. Gutfleisch, *IEEE Trans. Magn.* **2008**, *44*, 2993.

[10] N. Tiwari, S. Mishra, S. Sarkar, S. Talapatra, M. Palit, M. Paliwal, A. K. Singh, C. S. Tiwary, *J. Mater. Chem. C Mater.* **2025**, *13*, 10789.

[11] T. Krenke, E. Duman, M. Acet, E. F. Wassermann, X. Moya, L. Mañosa, A. Planes, E. Suard, B. Ouladdiaf, *Phys. Rev. B Condens. Matter Mater. Phys.* **2007**, *75*, DOI 10.1103/PhysRevB.75.104414.

[12] Y. Zhang, J. Bai, K. Guo, D. Liu, J. Gu, N. Morley, Q. Ma, Q. Gao, Y. Zhang, C. Esling, X. Zhao, L. Zuo, *J. Alloys Compd.* **2024**, *979*, DOI 10.1016/j.jallcom.2024.173593.

[13] T. Krenke, E. Duman, M. Acet, E. F. Wassermann, X. Moya, L. Manosa, A. Planes, *Nat. Mater.* **2005**, *4*, 450.

[14] I. Titov, M. Acet, M. Farle, D. González-Alonso, L. Mañosa, A. Planes, T. Krenke, *J. Appl. Phys.* **2012**, *112*, DOI 10.1063/1.4757425.

[15] F. Zhang, P. Feng, A. Kiecana, Z. Wu, Z. Bai, W. Li, H. Chen, W. Yin, X. W. Yan, F. Ma, N. van Dijk, E. Brück, Y. Ren, *Adv. Funct. Mater.* **2024**, *34*, DOI 10.1002/adfm.202409270.

[16] K. Ahn, *J. Alloys Compd.* **2024**, *978*, DOI 10.1016/j.jallcom.2023.173378.

[17] K. A. Gschneidner, A. Pecharsky, K. W. Dennis, *Some Observations on the Gd-Fich Side of the Gd-C System*, **1997**.

[18] T. V. Jayaraman, L. Boone, J. E. Shield, *J. Magn. Magn. Mater.* **2014**, *363*, 201.

[19] R. Bjørk, C. R. H. Bahl, M. Katter, *J. Magn. Magn. Mater.* **2010**, *322*, 3882.

[20] A. Gràcia-Condal, A. Planes, L. Mañosa, Z. Wei, J. Guo, D. Soto-Parra, J. Liu, *Phys. Rev. Mater.* **2022**, *6*, DOI 10.1103/PhysRevMaterials.6.084403.

[21] G. F. Dong, Z. Y. Gao, X. L. Zhang, W. Cai, J. H. Sui, *J. Mater. Sci.* **2011**, *46*, 4562.

[22] J. Guo, M. Zhong, W. Zhou, Y. Zhang, Z. Wu, Y. Li, J. Zhang, Y. Liu, H. Yang, *Materials* **2021**, *14*, DOI 10.3390/ma14092339.

[23] Y. Feng, J. H. Sui, Z. Y. Gao, J. Zhang, W. Cai, *Materials Science and Engineering: A* **2009**, *507*, 174.

[24] J. Y. Law, V. Franco, *J. Mater. Res.* **2023**, *38*, 37.

[25] J. Y. Law, V. Franco, *APL Mater.* **2021**, *9*, DOI 10.1063/5.0058388.

[26] L. Han, S. Zhu, Z. Rao, C. Scheu, D. Ponge, A. Ludwig, H. Zhang, O. Gutfleisch, H. Hahn, Z. Li, D. Raabe, *Nat. Rev. Mater.* **2024**, *9*, 846.

[27] M. C. Gao, D. B. Miracle, D. Maurice, X. Yan, Y. Zhang, J. A. Hawk, *J. Mater. Res.* **2018**, *33*, 3138.
[28] D. B. Miracle, O. N. Senkov, *Acta Mater.* **2017**, *122*, 448.
[29] Y. Yuan, Y. Wu, X. Tong, H. Zhang, H. Wang, X. J. Liu, L. Ma, H. L. Suo, Z. P. Lu, *Acta Mater.* **2017**, *125*, 481.
[30] C. W. Bale, E. Bélisle, P. Chartrand, S. A. Decterov, G. Eriksson, A. E. Gheribi, K. Hack, I. H. Jung, Y. B. Kang, J. Melançon, A. D. Pelton, S. Petersen, C. Robelin, J. Sangster, P. Spencer, M. A. Van Ende, *CALPHAD* **2016**, *54*, 35.
[31] C. Du, L. Hu, Q. Pan, K. Chen, P. Zhou, G. Wang, *Materials Science and Engineering: A* **2022**, *832*, DOI 10.1016/j.msea.2021.142413.
[32] H. E. Stanley, *Introduction to Phase Transitions and Critical Phenomena*, **n.d.**
[33] A. Arrott, J. E. Noakes, **1967**, *19*.
[34] B. K. Banerjee, *Physics Letters* **1964**, *12*, 15.
[35] Y. Zhang, P. Xu, J. Zhu, S. Yan, J. Zhang, L. Li, *Materials Today Physics* **2023**, *32*, DOI 10.1016/j.mtphys.2023.101031.
[36] D. D. Belyea, M. S. Lucas, E. Michel, J. Horwath, C. W. Miller, *Sci. Rep.* **2015**, *5*, DOI 10.1038/srep15755.
[37] S. M. Na, P. K. Lambert, H. Kim, J. Paglione, N. J. Jones, *AIP Adv.* **2019**, *9*, DOI 10.1063/1.5079394.
[38] K. Sarlar, A. Tekgül, I. Kucuk, *Current Applied Physics* **2020**, *20*, 18.
[39] J. Y. Law, L. M. Moreno-Ramírez, Á. Díaz-García, A. Martín-Cid, S. Kobayashi, S. Kawaguchi, T. Nakamura, V. Franco, *J. Alloys Compd.* **2021**, *855*, DOI 10.1016/j.jallcom.2020.157424.
[40] M. S. Lucas, D. Belyea, C. Bauer, N. Bryant, E. Michel, Z. Turgut, S. O. Leontsev, J. Horwath, S. L. Semiatin, M. E. McHenry, C. W. Miller, in *J. Appl. Phys.*, **2013**.
[41] J. Y. Law, Á. Díaz-García, L. M. Moreno-Ramírez, V. Franco, *Acta Mater.* **2021**, *212*, DOI 10.1016/j.actamat.2021.116931.
[42] A. Perrin, M. Sorescu, M. T. Burton, D. E. Laughlin, M. McHenry, *JOM* **2017**, *69*, 2125.
[43] J. Harris, Z. Leong, P. Gong, J. Cornide, C. Pughe, T. Hansen, A. Quintana-Nedelcos, R. Rowan-Robinson, U. Dahlborg, M. Calvo-Dahlborg, R. Goodall, M. Rainforth, N. Morley, *J. Phys. D Appl. Phys.* **2021**, *54*, DOI 10.1088/1361-6463/ac1139.
[44] M. Kurniawan, A. Perrin, P. Xu, V. Keylin, M. McHenry, *IEEE Magn. Lett.* **2016**, *7*, DOI 10.1109/LMAG.2016.2592462.
[45] R. F. Zhao, B. Ren, G. P. Zhang, Z. X. Liu, J. jian Zhang, *J. Magn. Magn. Mater.* **2018**, *468*, 14.
[46] B. Cantor, I. T. H. Chang, P. Knight, A. J. B. Vincent, *Materials Science and Engineering: A* **2004**, *375–377*, 213.
[47] J. W. Yeh, S. K. Chen, S. J. Lin, J. Y. Gan, T. S. Chin, T. T. Shun, C. H. Tsau, S. Y. Chang, *Adv. Eng. Mater.* **2004**, *6*, 299.
[48] V. Chaudhary, V. Soni, B. Gwalani, R. V. Ramanujan, R. Banerjee, *Scr. Mater.* **2020**, *182*, 99.
[49] S. Huang, Á. Vida, A. Heczel, E. Holmström, L. Vitos, *JOM* **2017**, *69*, 2107.
[50] H. B. Tran, H. Li, *J. Mater. Chem. C Mater.* **2025**, *13*, 11393.
[51] A. V. Ruban, O. E. Peil, *Phys. Rev. B* **2018**, *97*, DOI 10.1103/PhysRevB.97.174426.
[52] H. Guan, S. Huang, J. Ding, F. Tian, Q. Xu, J. Zhao, *Acta Mater.* **2020**, *187*, 122.
[53] Z. Rao, B. Dutta, F. Körmann, D. Ponge, L. Li, J. He, L. Stephenson, L. Schäfer, K. Skokov, O. Gutfleisch, D. Raabe, Z. Li, *Phys. Rev. Mater.* **2020**, *4*, DOI 10.1103/PhysRevMaterials.4.014402.
[54] S. Huang, E. Holmström, O. Eriksson, L. Vitos, *Intermetallics (Barking).* **2018**, *95*, 80.

[55] B. Uthaman, G. R. Raji, S. Thomas, K. G. Suresh, M. Raama Varma, *Intermetallics (Barking).* **2019**, *115*, DOI 10.1016/j.intermet.2019.106629.
[56] S. E. Muthu, G. Kalaiselvan, R. J. Joseyphus, *Materials Science and Engineering: B* **2025**, *313*, DOI 10.1016/j.mseb.2024.117880.
[57] A. García, N. Papior, A. Akhtar, E. Artacho, V. Blum, E. Bosoni, P. Brandimarte, M. Brandbyge, J. I. Cerdá, F. Corsetti, R. Cuadrado, V. Dikan, J. Ferrer, J. Gale, P. García-Fernández, V. M. García-Suárez, S. García, G. Huhs, S. Illera, R. Korytár, P. Koval, I. Lebedeva, L. Lin, P. López-Tarifa, S. G. Mayo, S. Mohr, P. Ordejón, A. Postnikov, Y. Pouillon, M. Pruneda, R. Robles, D. Sánchez-Portal, J. M. Soler, R. Ullah, V. W. Z. Yu, J. Junquera, *Journal of Chemical Physics* **2020**, *152*, DOI 10.1063/5.0005077.
[58] J. M. Soler, E. Artacho, J. D. Gale, A. García, J. Junquera, P. Ordejón, D. Sánchez-Portal, *The SIESTA Method for Ab Initio Order-N Materials Simulation*, **2002**.
[59] D. M. Bylander, L. Kleinman, *Phys. Rev. Lett.* **1982**, *48*, 1425.
[60] J. P. Perdew, K. Burke, M. Ernzerhof, *Generalized Gradient Approximation Made Simple*, **1996**.
[61] H. J. Monkhorst, J. D. Pack, *Special Points for Brillonin-Zone Integrations**, **1976**.
[62] X. He, N. Helbig, M. J. Verstraete, E. Bousquet, *Comput. Phys. Commun.* **2021**, *264*, 107938.
[63] R. F. L. Evans, W. J. Fan, P. Chureemart, T. A. Ostler, M. O. A. Ellis, R. W. Chantrell, *Journal of Physics Condensed Matter* **2014**, *26*, DOI 10.1088/0953-8984/26/10/103202.
[64] J. D. Alzate-Cardona, D. Sabogal-Suárez, R. F. L. Evans, E. Restrepo-Parra, *Journal of Physics Condensed Matter* **2019**, *31*, DOI 10.1088/1361-648X/aaf852.
[65] R. F. L. Evans, in *Handbook of Materials Modeling*, Springer International Publishing, **2018**, pp. 1–23.

The present study highlights the exceptional compositional flexibility of rare-earth-free high-entropy alloys in achieving a wide operational temperature range. By tuning only a single constituent element, namely Cu, the investigated alloys exhibit magnetocaloric performance from cryogenic to near-room-temperature conditions, demonstrating their significant potential for sustainable and efficient magnetic refrigeration applications.

**Composition-Driven High-Entropy Alloys with Enhanced Magnetocaloric Properties**

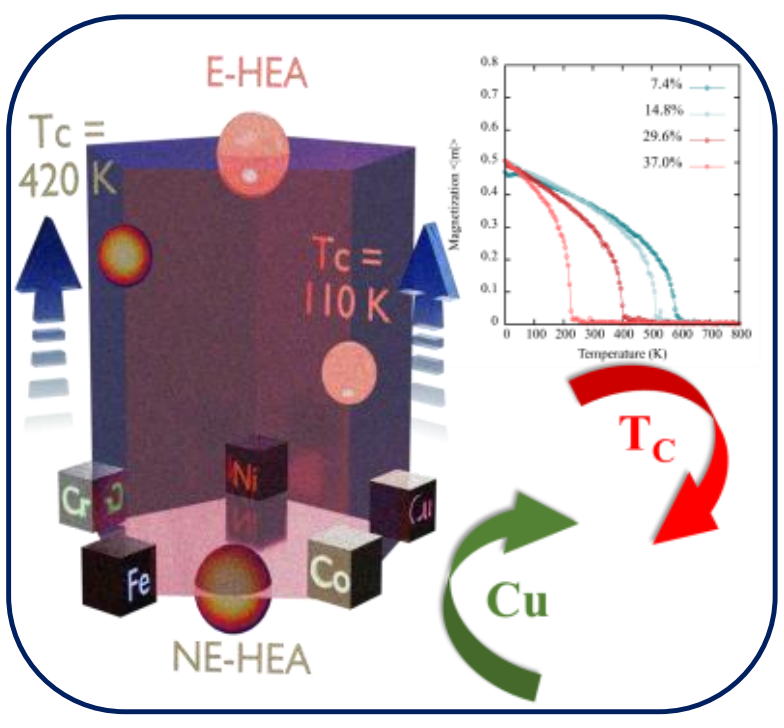

# Supporting Information

**Composition-Driven High-Entropy Alloys with Enhanced Magnetocaloric Properties**

*Nishant Tiwari [a, bΨ], Juan Rafael Gomez Quispe [cΨ], Noorbasha Bhavani Sai[d], Saikat Talapatra[e], Pedro Alves Da Silva Autreto*[c], Varun Chaudhary*[b] and Chandra Sekhar Tiwary*[a, d]*

Nishant Tiwari, Chandra Sekhar Tiwary
Department of Metallurgical and Materials Engineering, Indian Institute of Technology Kharagpur, West Bengal 721302, India
E-mail: Chandra.tiwary@metal.iitkgp.ac.in

Nishant Tiwari, Varun Chaudhary
Department of Mechanical Engineering, Chalmers University of Technology, Gothenburg SE-41296, Sweden
Email: varunc@chalmers.se

Juan Rafael Gomez Quispe, Pedro Alves Da Silva Autreto
Center for Natural and Human Sciences (CCNH), Federal University of ABC Rua Santa Adélia 166, Santo André 09210-170, Brazil
Email: pedro.autreto@uafbc.edu.br

Noorbasha Bhavani Sai
School of Nano Science and Technology, Indian Institute of Technology, Kharagpur, West Bengal 721302, India

Saikat Talapatra
School of Physics and Applied Physics, Southern Illinois University, Carbondale, IL 62901, USA

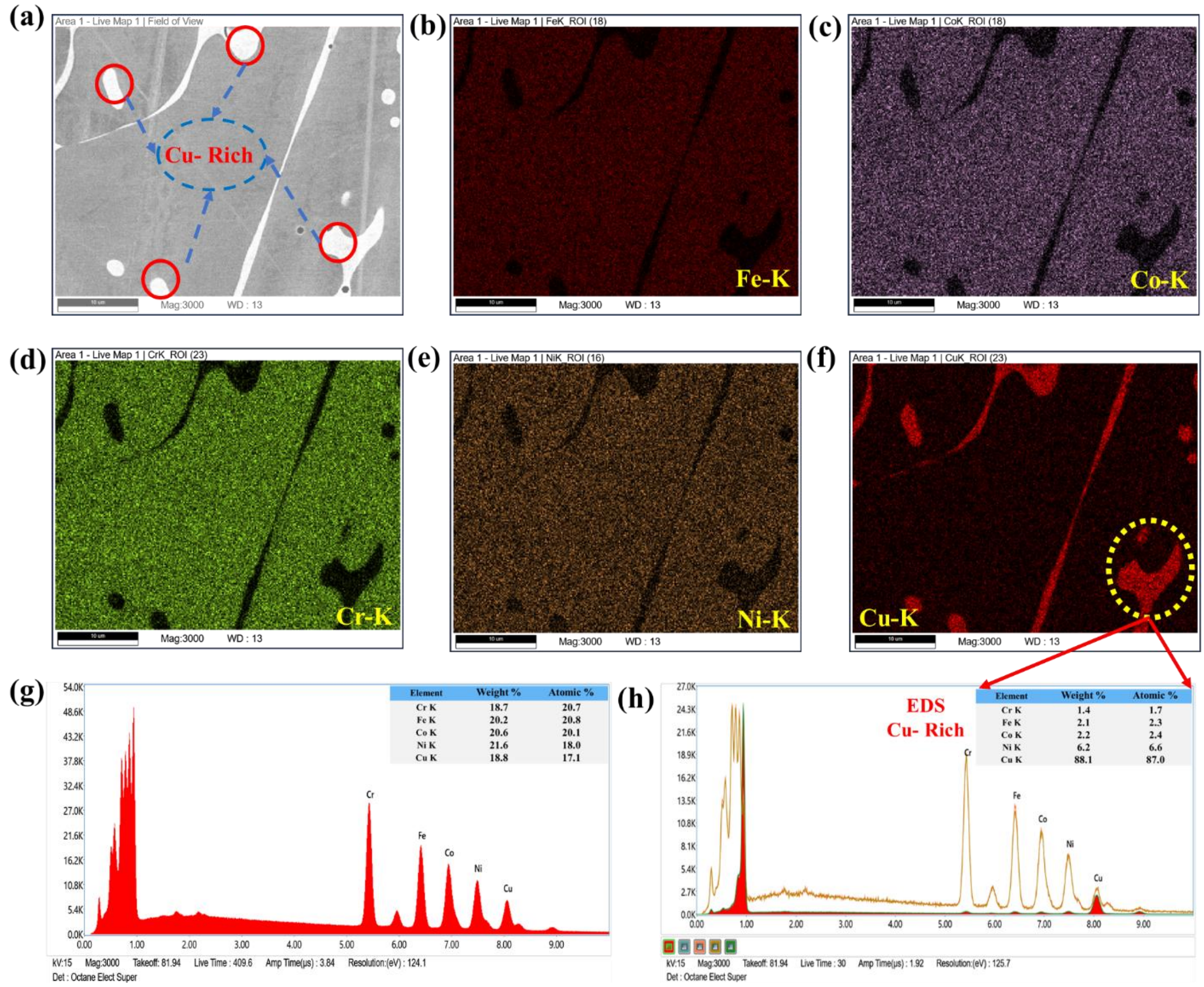

**Figure S1:** (a) BSE image of E-HEA showing Cu rich precipitates. (b-f) Elemental mapping of Fe, Co, Cr, Ni, and Cu elements. (g) EDS mapping of E-HEA. (h) EDS mapping of Cu rich phase.

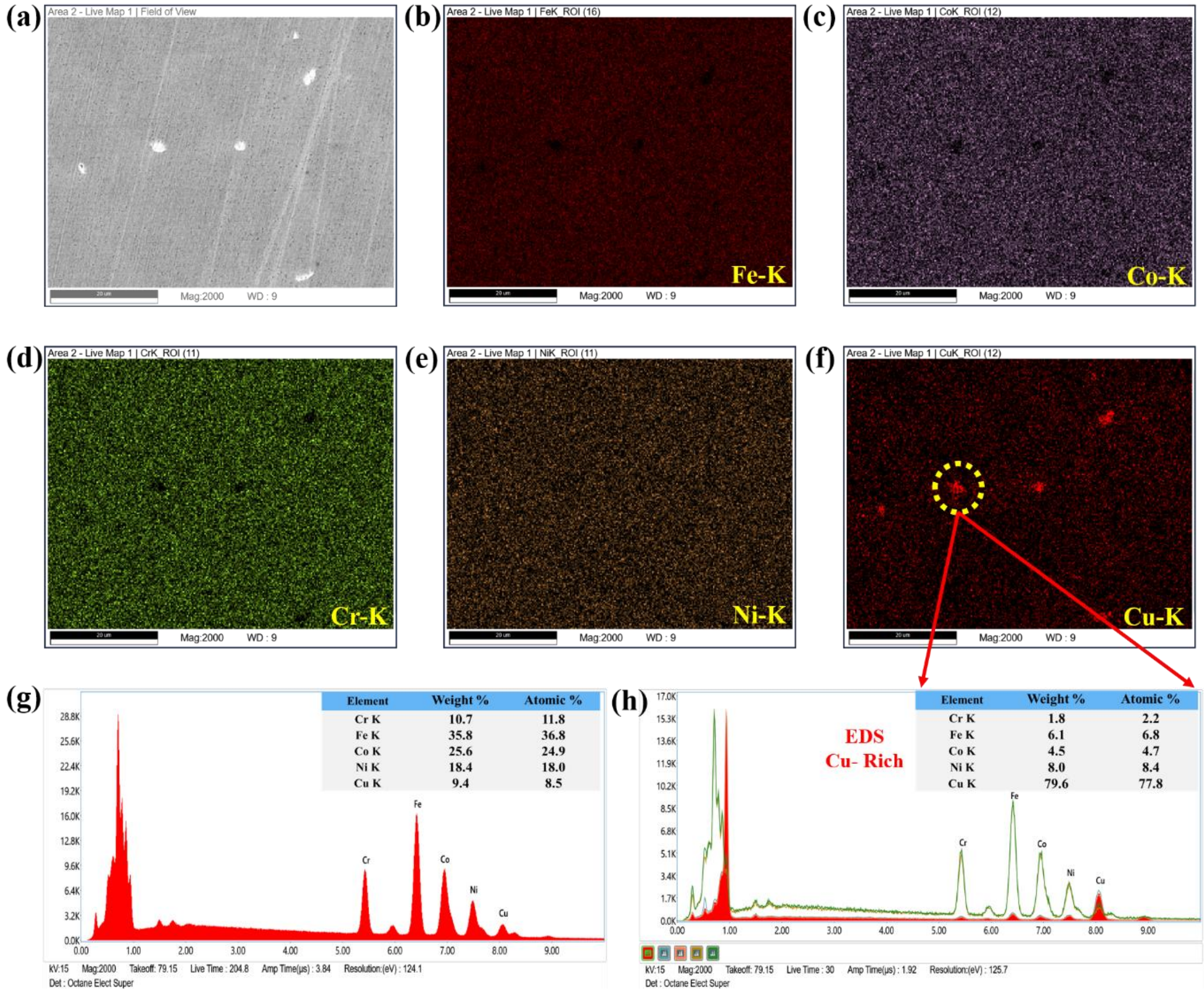

**Figure S2: (a)** BSE image of E-HEA showing Cu-rich precipitates. **(b-f)** Elemental mapping of Fe, Co, Cr, Ni, and Cu. **(g)** EDS mapping of E-HEA. **(h)** EDS mapping of the Cu-rich phase.